\documentclass[pre,twocolumn,showpacs]{revtex4}

\usepackage{graphicx}
\usepackage{dcolumn}
\usepackage{bm}

\begin{document}
\title{Anomalous diffusion. A competition between the very large jumps\\ in physical
and operational times}

\author{Aleksander Stanislavsky}
\affiliation{Institute of Radio Astronomy, 4 Chervonopraporna St., 61002 Kharkov,
Ukraine} \email{alexstan@ri.kharkov.ua}
\author{Karina Weron}%
\affiliation{Institute of Physics,  Wroc{\l}aw University of Technology, Wyb.
Wyspia$\acute{n}$kiego 27, 50-370 Wroc{\l}aw, Poland}%
\email{karina.weron@pwr.wroc.pl}
\date{\today}

\begin{abstract}
In this paper we analyze a coupling between the very large jumps
in physical and operational times as applied to anomalous
diffusion. The approach is based on subordination of a skewed
L\'evy-stable process by its inverse to get two types of
operational time -- the spent and the residual waiting time,
respectively. The studied processes have different properties
which display both subdiffusive and superdiffusive features of
anomalous diffusion underlying the two-power-law relaxation
patterns.
\end{abstract}

\pacs{05.40.Fb, 02.50.Ey, 05.10.Gg}

\maketitle

\section{Introduction}
The Continuous Time Random Walk (CTRW) formalism is a very
powerful stochastic approach to model physical processes
demonstrating anomalous diffusion and slow, power-law, relaxation.
It describes random walks in space and time by means of iid
(independent and identically distributed) couples of space and
time random steps $(R_i,T_i)$. The simplest, decoupled CTRW
considers independent time and space steps. This model involves
stable distributions, and it shows various anomalous behaviors
like subdiffusion (diffusion slower than normal one),
Mittag-Leffler relaxation and fractional diffusive equations
\cite{mk04,stan04,magwer06}. A more complex CTRW model accounts
for coupling between time and space steps. The coupled CTRWs were
considered in the context of anomalous diffusion and
non-exponential relaxation \cite{bkms04,meersca06,stan03}. In this
case the anomalous diffusion evolution is much richer. Sub- and
superdiffusion (faster than normal) can  be modeled. However, the
analysis is rather exotic for the research, and it is in progress.
Recently, the anomalous subdiffusive behavior attracts a great
attention in modeling of subdiffusion in space-time-dependent
force fields beyond the fractional Fokker-Planck equation
\cite{wmw08,mwk08}. This approach uses the Langevin-type dynamics
with subordination techniques, where the force depends on a
compound subordinator. It is coupled because of a L\'evy-stable
process directed by its inverse. The fractional two-power-law
relaxation can be also described in the framework of coupled CTRWs
based on subordination of a stochastic process with the
heavy-tailed distribution of the waiting times by its inverse
\cite{jwt2008,wjmwt2010}. Although the papers have a different
physical background, they intersect into the application of the
coupling between the L\'evy-stable process and its inverse.
Undoubtedly, this new random process (subordinator) is of an
essential interest for understanding of the anomalous relaxation
phenomena and was investigated unsufficiently yet. In this paper
we are going to make up the deficiency.

\section{Coupling between the very large jumps in physical and operational times}\label{par2}
The probability density of the position vector ${\bf r}_t={\bf
B}_{S_t}$  (where ${\bf B}_\tau$ is the standard Brownian motion)
can be found from a weighted integration of the joint probability
density of the couple $({\bf R}_\tau, T_\tau)$ over the internal
time parameter $\tau$ by subordination. The stochastic time
evolution $T_\tau$ and its (left) inverse process $S_t$ permits
one to underestimate or overestimate the physical time $t$.

The sum of iid heavy-tailed random variables $T_i$
\begin{equation}
{\rm Pr}(T_i\geq t)\sim\Bigg(\frac{t}{t_0}\Bigg)^{-\alpha}\quad{\rm as}
\quad t\to\infty\,,\label{eq0}
\end{equation}
$0<\alpha<1$, $t_0>0$ converges to a stable random variable in
distribution as the number of summands tends to infinity. Let
$U_n=\sum_{i=0}^nT_i$ with $T_0=0$. The counting process
$N_t=\max\{n\in {\bf N}\,|\,U_n\leq t\}$ is inverse to $U_n$ which
can be defined equivalently as the process satisfying
\begin{equation}
U_{N_t}< t < U_{N_t+1}\quad {\rm for}\quad t> 0\,,\label{eq1}
\end{equation}
what follows directly from its definition. In fact, the two
processes $U_{N_t}$ and $U_{N_t+1}$ correspond to underestimating
and overestimating the real time $t$ from the random time steps $T_i$ of
the CTRWs.

In terminology of the Feller's book \cite{feller} the variable
$Z_t=U_{N_t+1}-t$ is the residual waiting time (life-time) at the
epoch $t$, and $Y_t=t-U_{N_t}$ is the spent waiting time (age of
the object that is alive at time $t$). The importance of these
variables can be explained by one remarkable property. For
$t\to\infty$ the variables $Y_t$ and $Z_t$ have a common proper
limit distribution only if their probability distributions $F(y)$
and $F(z)$ have finite expectations. However, if the distribution
$F(x)$ satisfies $1-F(x)=x^{-\alpha}L(x)$, where $0<\alpha<1$ and
$L(xt)/L(t)\to 1$ as $x\to\infty$, then according to \cite{dyn61},
the probability density function (pdf) of the normalized variable
$Y_t/t$ is given by the generalized arc sine law
\begin{equation}
p_\alpha(x)=\frac{\sin(\pi\alpha)}{\pi}\,x^{-\alpha}(1-x)^{\alpha-1}\,,\label{eq2}
\end{equation}
while $Z_t/t$ obeys
\begin{equation}
q_\alpha(x)=\frac{\sin(\pi\alpha)}{\pi}\,x^{-\alpha}(1+x)^{-1}\,.\label{eq3}
\end{equation}
Since $\Sigma_{N_t}=t-Y_t$ and $\Sigma_{N_t+1}=Z_t+t$, the
distributions of $\Sigma_{N_t}/t$ and $\Sigma_{N_t+1}/t$ can be obtained
from Eqs. (\ref{eq2}) and (\ref{eq3}) by a simple change of
variables $1-x=y$ and $1+x=z$, respectively.

We now return to the processes $U_{N_t}$ and $U_{N_t+1}$
introduced above. Recall that $T_i$ are iid positive random
variables with a long-tailed distribution (\ref{eq0}). In this
case $U_{N_t}/t$ tends in distribution
($\stackrel{d}{\rightarrow}$) in the long-time limit to random
variable $Y$ with density
\begin{equation}
p^Y(x)=\frac{\sin(\pi\alpha)}{\pi}\,x^{\alpha-1}(1-x)^{-\alpha}\,,\quad
0<x<1 \label{eq4}
\end{equation}
and $U_{N_t+1}/t\stackrel{d}{\rightarrow}Z$ with the pdf equal to
\begin{equation}
p^Z(x)=\frac{\sin(\pi\alpha)}{\pi}\,x^{-1}(x-1)^{-\alpha}\,,\quad
x>1.\label{eq5}
\end{equation}
The functions $p^Y(x)$ and $p^Z(x)$ correspond to special cases of
the well-known beta density. It should be noticed that the density
$p^Y(x)$ concentrates near 0 and 1, whereas $p^Z(x)$ does near 1.
Near 1 both tend to infinity. This means that in the long-time
limit the most probable values for $U_{N_t}$ occur near 0 and 1,
while for $U_{N_t+1}$ they tend to be situated near 1.

As a consequence, the random variable $Y$ has finite moments of
any order. They can be calculated directly from the density
(\ref{eq4}) and take the form
\begin{eqnarray}
\langle Y\rangle&=&\alpha,\,\langle
Y^2\rangle=\frac{\alpha(1+\alpha)}{2}\,,\,\dots\,,\,\nonumber\\
\langle Y^n\rangle&=&\frac{\alpha(1+\alpha)\dots(\alpha+n-1)}{n!}\,,\nonumber
\end{eqnarray}
where $n\in{\bf N}$, while even the first moment of $Z$ diverges.
The divergence of $U_{N_t+1}$ results from the long-tail property
(\ref{eq0}) of the time steps $T_i$ so that $\langle
T_i\rangle=\infty$, yielding too long overshot above $t$.

\begin{figure}
\centering
\includegraphics[width=8.6 cm]{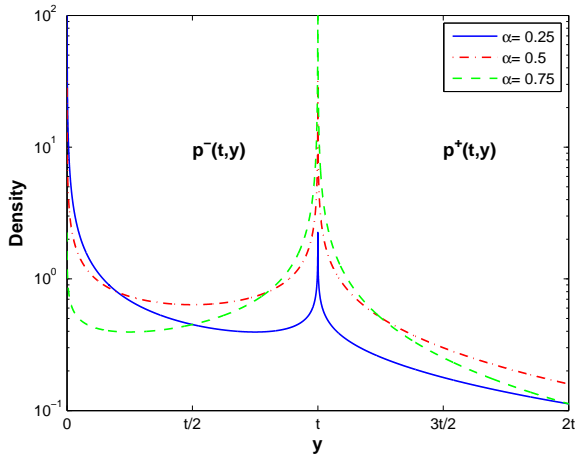}
\caption{(Color online) The probability density $p^-(y)$ with
support on $0<y<t$ and the density $p^+(y)$ with support on $y>t$
for different values of the index $\alpha$.}\label{fig1}
\end{figure}

The nonequality (\ref{eq1}) can also be represented in a schematic
picture of time steps as
$T^-_\tau(\Delta\tau)=U_{[\tau/\Delta\tau]}$ and
$T_\tau(\Delta\tau)=U_{[\tau/\Delta\tau]+1}\,$, where
$T^-_\tau(\Delta\tau)=\lim_{\epsilon\downarrow
0}T_{\tau-\epsilon}(\Delta\tau)$ is the left-limit process, and
$[x]$ indicates the integer part of the real number $x$ so that
$[x]\leq x <[x]+1$. The inverse process of $T^-_\tau(\Delta\tau)$
and $T_\tau(\Delta\tau)$ is $S_t(\Delta\tau)=\inf\{\tau
\geq0\,|\,T_\tau(\Delta\tau)>t\}$ or equivalently
$S_t(\Delta\tau)=\Delta\tau\,N_t$. Therefore, in the limit
$\Delta\tau\to 0$ the processes $U_{N_t+1}$ and $U_{N_t}$ can be
expressed through the stochastic process $T_\tau$ and its left
limit subordinated by their inverse:
\begin{displaymath}
U_{N_t}\stackrel{d}{\rightarrow}T^-_{S_t}\quad{\rm and}\quad
U_{N_t+1}\stackrel{d}{\rightarrow}T_{S_t}\,.
\end{displaymath}
The passage from the discrete process $T_i$ to the continuous one
$T_\tau$ allows one to reformulate the unequality (\ref{eq1}) as
\begin{equation}
T^-_{S_t}< t < T_{S_t}\quad {\rm for}\quad t>0\,.\,,\label{eq6}
\end{equation}
underestimating or overestimating the real time $t$. From Theorem
1.13 in \cite{ber97} the joint probability density $p(y,z)$ of
$T^-_{S_t}$ and $T_{S_t}$ with $0\leq T^-_{S_t} \leq t<T_{S_t}$
takes the form
\begin{equation}
p(y,z)=\frac{\alpha\sin(\pi\alpha)}{\pi}\,
y^{\alpha-1}(z-y)^{-1-\alpha}\label{eq7}
\end{equation}
for $0\leq y\leq t <z$. After integrating (\ref{eq7}) with respect
to $z$ in the limits $[t\ ,\infty[$ (or with respect to $y$ in the
limits $[0\ ,t]$) we obtain the densities of $T^-_{S_t}$ and
$T_{S_t}$, respectively
\begin{eqnarray}
p^-(t,y)&=&
\frac{\sin\pi\alpha}{\pi}\,y^{\alpha-1}(t-y)^{-\alpha}\,,\quad
0<y<t\,,\label{eq8}\\
p^+(t,z)&=&\frac{\sin\pi\alpha}{\pi}\,z^{-1}\,t^\alpha
(z-t)^{-\alpha}\,,\quad z>t\,,\label{eq9}
\end{eqnarray}
valid for any time $t>0$ (see Fig.~\ref{fig1}). The moments of
$T^-_{S_t}$ and $T_{S_t}$ can be calculated directly from the
moments of $Y$ and $Z$ by using relations
\begin{displaymath}
T^-_{S_t}\stackrel{d}{=}tY\quad{\rm and}\quad
T_{S_t}\stackrel{d}{=}tZ\,,
\end{displaymath}
where $\stackrel{d}{=}$ means the equality in distribution. Thus,
the process $T^-_{S_t}$ has finite moments of any order, while
$T_{S_t}$ gives us even no finite the first moment. The overshot
of $T_{S_t}>t$ is too long also in the limit formulation. Notice
that $p^+(t,y)=y^{-2}p^-(t^{-1},y^{-1})$. At this point we should
mention that compound subordinators, and in particular the
subordination by an inverse L\'evy-stable process via a
L\'evy-stable process, were considered already in \cite{huil00}.
However, the construction of compound subordinators has been based
on the statistically independent stochastic processes. This leads
to quite different results in comparison with ours. In our
construction of the compound subordinators $T^-_{S_t}$ and
$T_{S_t}$ the processes $U_t$ and $S_t$ are clearly coupled.

\section{Anomalous diffusion with under- and overshooting subordination}
\label{par3} According to \cite{wjmwt2010}, the widely observed
fractional two-power relaxation dependencies
\begin{equation}
\chi(\omega)\sim\left(i\omega/\omega_p\right)^{n-1}\quad {\rm
for}\quad\omega\gg\omega_p \label{eqi}
\end{equation}
and
\begin{equation}
\Delta\chi(\omega)\sim\left(i\omega/\omega_p\right)^m\quad {\rm
for}\quad\omega\ll\omega_p \label{eqii}
\end{equation}
of the complex susceptibility
$\chi(\omega)=\chi'(\omega)-i\chi''(\omega)$, where
$\Delta\chi(\omega)=\chi(0)-\chi(\omega)$, the exponent $n$ and
$m$ fall in the range $(0,1)$, and $\omega_p$ denotes the loss
peak frequency, are closely connected with the under- and
overshooting subordination
\begin{displaymath}
Z^U_{\alpha,\gamma}(t)< S_\alpha(t) < Z^O_{\alpha,\gamma}(t)\quad
{\rm for}\quad t> 0\,,
\end{displaymath}
where $Z^U_{\alpha,\gamma}(t)=X^U_\gamma[S_\alpha(t)]$,
$Z^O_{\alpha,\gamma}(t)=X^O_\gamma[S_\alpha(t)]$. Here the
processes $X^U_\gamma(t)$ and $X^O_\gamma(t)$ are nothing else as
$T^-_{S_t}$ and $T_{S_t}$ with the index $\gamma$. They are
subordinated by an independent inverse $\alpha$-stable process
$S_\alpha(t)$ forming the compound subordinators
$Z^U_{\alpha,\gamma}(t)$ and $Z^O_{\alpha,\gamma}(t)$,
respectively. The approach enlarges the class of diffusive
scenarios in the framework of the CTRWs. This new type of coupled
CTRWs follows from the clustering-jump random walks idea
\cite{vlad}. As it has been rigorously proved \cite{urlew05}, the
clustering with finite-mean-value cluster sizes leads to the
classical decoupled CTRW models, but assuming a heavy-tailed
cluster-size distribution with the tail exponent $0<\gamma<1$, the
coupling between jumps and interjump times tends to the compound
operational times $Z^U_{\alpha,\gamma}(t)$ and
$Z^O_{\alpha,\gamma}(t)$ as under- and overshooting subordinators,
respectively.

The overshooting subordinator yields the anomalous diffusion
scenario leading to the well-known Havriliak-Negami relaxation
pattern \cite{J}, and the undershooting subordinator leads to a
new relaxation law given by the generalized Mittag-Leffler
relaxation function \cite{jwt2008,wjmwt2010}. These results are in
agreement with the idea of a superposition of the classical
(exponential) Debye relaxations. Thus, the stochastic mechanism
underlying the anomalous relaxation is quite clear, but the
corresponding diffusion analysis requires some additional clarity.
Let $B(t)$ be the parent process that is subordinated either by
$Z^U_{\alpha,\gamma}(t)$ or $Z^O_{\alpha,\gamma}(t)$. Then the
subordination relation, expressed by means of a mixture of pdf's,
takes the form
\begin{equation}
p^{\,r}(x,t)=\int_0^\infty\int_0^\infty
p^B(x,y)\,p^\pm(y,\tau)\,p^S(\tau,t)\,dy\,d\tau\,,\label{eq10}
\end{equation}
where $p^{\,r}(x,t)$ is the probability density of the
subordinated process $B[Z^U_{\alpha,\gamma}(t)]$ (or
$B[Z^O_{\alpha,\gamma}(t)]$) with respect to the coordinate $x$
and time $t$, $p^B(x,\tau)$ the probability density of the parent
process, $p^\pm(y,\tau)$ the probability density of $T^-_{S_t}$
and $T_{S_t}$ respectively, and $p^S(\tau,t)$ the probability
density of $S(t)$. Recall that for the subdiffusion $B[S(t)]$, by
taking the Laplace transform from the corresponding subordination
relation, we can derive the celebrated fractional Fokker-Planck
equation \cite{km2000}. It is therefore reasonable to ask is it
possible to find a diffusion equation corresponding to relation
(\ref{eq10}). In the Laplace space
\begin{displaymath}
\bar{f}(u)=\int^\infty_0e^{-ut}\,f(t)\,dt
\end{displaymath}
we obtain
\begin{equation}
\bar{p}^{\,r}(x,u)=u^{\alpha-1}\int_1^\infty
\bar{p}^B(x,u^\alpha/z)\,p^+_0(z)\,\frac{dz}{z}\label{eq11}\\
\end{equation}
with $p^+_0(z)=\sin(\pi\gamma)\,z^{-1}(z-1)^{-\gamma}/\pi$ for
$z>1$, as well
\begin{equation}
\bar{p}^{\,r}(x,u)=u^{\alpha-1}\int_0^1
\bar{p}^B(x,u^\alpha/z)\,p^-_0(z)\,\frac{dz}{z}\label{eq12}
\end{equation}
with $p^-_0(z)=\sin(\pi\gamma)\,z^{\gamma-1}(1-z)^{-\gamma}/\pi$
for $0<z<1$. The Laplace image of the pdf of the subordinated
process $B[S(t)]$ can be simply expressed in terms of an algebraic
form with the Laplace image of the parent process pdf. This allows
one to get the fractional Fokker-Planck equation driving the
spatio-temperal evolution of the propagator of the anomalous
diffusion underlying the Mittag-Leffler relaxation
\cite{magwer06,stan03,km2000}. However, expressions (\ref{eq11})
and (\ref{eq12}) are not similar to the latter. They have an
integral form. Nevertheless, derivation of the corresponding
Fokker-Planck equation is also possible.

If we take the Laplace transform with respect to $t$ and the
Fourier transform with respect to $x$ for $p^{\,r}(x,t)$ in Eq.
(\ref{eq10}), the Fourier-Laplace (FL) image reads
\begin{eqnarray}
&&\textbf{FL }(p^{\,r})(k,s)\nonumber\\ &=&s^{\alpha-1}\int_0^\infty\int_0^\infty
e^{-\psi(k)y}\,p^\pm(y,\tau)\,e^{-\tau s^\alpha}\,dy\,d\tau\,,\label{eq12a}
\end{eqnarray}
where $\psi(k)$ is the log-Fourier transform of the parent process
pdf $p^B(x,y)$. Consider the case of $p^-(y,\tau)$. After changing
variables $y=z\tau$ we take the integral
\begin{displaymath}
\int^\infty_0e^{-\tau(s^\alpha+\psi(k)z)}\,d\tau = \frac{1}{s^\alpha+\psi(k)z}\,.
\end{displaymath}
Next, the change of variables $t=z/(1-z)$ maps $[0\,,\,1]$ onto
$[0\,,\,\infty)$. This helps to derive
\begin{displaymath}
\textbf{FL }(p^{\,r})(k,s)=\frac{s^{\alpha-1}}{\Gamma(\gamma)
\Gamma(1-\gamma)}\int^\infty_0\frac{t^{\gamma-1}\,dt}{(s^\alpha+
\psi(k))t+s^\alpha}\,.
\end{displaymath}
The last expression can be easily calculated from the integral \cite{abr64}
\begin{displaymath}
\int^\infty_0\frac{t^{\gamma-1}}{t+1}\,dt=\Gamma(\gamma)\Gamma(1-\gamma)\,.
\end{displaymath}
The FL image of $p^{\,r}(x,t)$ with the undershooting directing
process $Z^U_{\alpha,\gamma}(t)=X^U_\gamma[S_\alpha(t)]$ is of the
form
\begin{equation}
\textbf{FL }(p^{\,r})(k,s)=\frac{s^{\alpha\gamma-1}}{(s^\alpha+
\psi(k))^\gamma}\,.\label{eq12b}
\end{equation}
Finally, we invert the Fourier and Lapace transforms to get the pseudo-differential equation
\begin{equation}
\Bigg[\frac{\partial^\alpha}{\partial t^\alpha}+L_{\rm
FP}(x)\Bigg]^\gamma\,p^{\,r}(x,t)=\delta(x)
\frac{t^{-\alpha\gamma}}{\Gamma(1-\alpha\gamma)} \,,\label{eq12c}
\end{equation}
where $L_{\rm FP}(x)$ is the Fokker-Planck operator, $\delta(x)$
the Dirac function, and $\partial^\alpha/\partial t^\alpha$
denotes the Riemann-Louiville derivative. The corresponding
Fokker-Planck equation can be obtained also in the case when the
overshooting directing process
$Z^O_{\alpha,\gamma}(t)=X^O_\gamma[S_\alpha(t)]$ is taken into
account. Unfortunately, the derivation is more complicated as we
present below.

In the case of $p^+(y,\tau)$, after the substitution $y=z\tau$, we
map $[1\,,\,\infty)$ onto $[0\,,\,1]$ by the change of variables
$z=1/x$. Then we obtain the corresponding FL image
\begin{displaymath}
\textbf{FL }(p^{\,r})(k,s)=\frac{s^{\alpha-1}}{\Gamma(\gamma)
\Gamma(1-\gamma)}\int^1_0\frac{x^{\gamma-1}\,(1-x)^{-\gamma}\,dt}
{s^\alpha+\psi(k)/x}\,.
\end{displaymath}
The mapping $t=x/(1-x)$ transforms the latter expression to the
form
\begin{eqnarray}
&&\textbf{FL }(p^{\,r})(k,s)\nonumber\\ &=&\frac{s^{\alpha-1}}
{\Gamma(\gamma)\Gamma(1-\gamma)}\int^\infty_0\frac{t^\gamma
\,dt}{(1+t)[(s^\alpha+\psi(k))t+\psi(k)]}\,.\nonumber
\end{eqnarray}
This integral can be calculated exactly:
\begin{displaymath}
\int^\infty_0\frac{t^\gamma}{(t+1)(at+b)}\,dt=\frac{\Gamma(\gamma)\Gamma(1-\gamma)}{(a-b)}
\Big[1-(b/a)^\gamma\Big]\,.
\end{displaymath}
As a result, the FL image of $p^{\,r}(x,t)$ with the directing
process $Z^O_{\alpha,\gamma}(t)=X^O_\gamma[S_\alpha(t)]$, can be
written as
\begin{equation}
\textbf{FL
}(p^{\,r})(k,s)=\frac{1}{s}\Bigg\{1-\left(\frac{\psi(k)}
{s^\alpha+\psi(k)}\right)^\gamma\Bigg\}\,.\label{eq12d}
\end{equation}
Now we invert the Fourier and Laplace transforms to get the
pseudo-differential equation
\begin{equation}
\Bigg[\frac{\partial^\alpha}{\partial t^\alpha}+L_{\rm FP}(x)
\Bigg]^\gamma\,p^{\,r}(x,t)=f_{\alpha,\gamma}(x,t)\,,\label{eq12e}
\end{equation}
where
\begin{displaymath}
f_{\alpha,\gamma}(x,t)=\Bigg{\{}\Bigg[\frac{\partial^\alpha}
{\partial t^\alpha}+L_{\rm FP}(x)\Bigg]^\gamma-\Bigg[L_{\rm
FP}(x)\Bigg]^\gamma\Bigg{\}}\,\delta(x)
\end{displaymath}
is a function depending on the probability density $p^B(x,y)$. The
exact form of $f_{\alpha,\gamma}(x,t)$ is quite different from the
right-side term of Eq.(\ref{eq12c}). In this connection it should
be pointed out the work \cite{kcct04}, where the derivation of a
fractional Fokker-Planck underlying the Havriliak-Negami type of
relaxation is based on the entirely phenomenological approach of
\cite{nr97}. However, the stochastic background leading to the
anomalous diffusion yielding the Havriliak-Negami pattern, has
remained behind these works. It should be noticed that
Eqs.(\ref{eq12c}) and (\ref{eq12e}) have been derived
independently in papers \cite{MMOCTRW,StuMath}.

\begin{figure}
\centering
\includegraphics[width=8.6 cm]{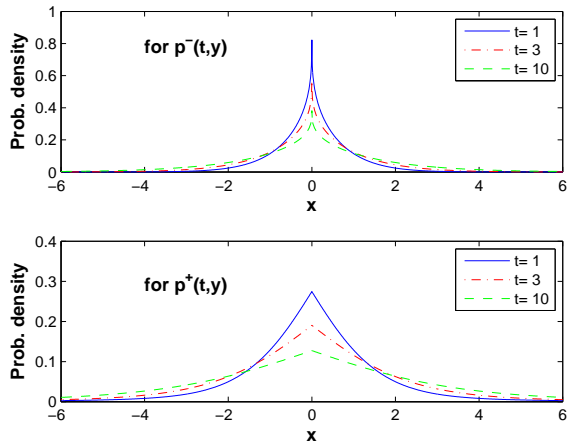}
\caption{(Color online) Propagator $p^r(x,t)$ of the under- and
overshooting anomalous diffusion with a constant potential,
$\alpha=2/3$ and $\gamma=2/3$, drawn for consecutive dimensionless
instances of time $t=1,3,10$. The cusp shape of the pdfs
appears.}\label{fig2}
\end{figure}

To calculate the moments of the processes
$B[Z^U_{\alpha,\gamma}(t)]$ and $B[Z^O_{\alpha,\gamma}(t)]$,
assume for simplicity, that the parent process $B$ is a
one-dimensional Brownian motion. Its moments are written as
\begin{eqnarray}
I_{2n}(t)&=&\frac{1}{\sqrt{4\pi
Dt}}\int^\infty_{-\infty}x^{2n}\,\exp\Bigg(-\frac{x^2}{4Dt}\Bigg)\,
dx\nonumber\\
&=&\frac{(2n)!}{n!}\,(Dt)^n\,,\nonumber
\end{eqnarray}
where $D$ is the diffusion coefficient. If the subordinator
$Z^U_{\alpha,\gamma}(t)$ governs the Brownian motion, then the
moment integral reads
\begin{eqnarray}
<x^{2n}>&=&\int^\infty_{-\infty}x^{2n}\,
p^{\,r}(x,t)\,dx\nonumber\\ &=&B_n\int^1_0z^n\,p^-_0(z)\,dz\int_0^\infty\tau^n\,
p^S(\tau,y)\,d\tau\nonumber\\  &=&\frac{
(2n)!}{n!}\,D^n\frac{(\gamma,n)}{n!}
\frac{t^{n\alpha}}{\Gamma(1+n\alpha)}\,,\label{eq13}
\end{eqnarray}
where $(\gamma,n)=\gamma(\gamma+1)(\gamma+2)\dots(\gamma+n-1)$ is
the Appell's symbol with $(\gamma,0)=1$. When another subordinator
$Z^O_{\alpha,\gamma}(t)$ is used, even the first moment of the
subordinated process $B[Z^O_{\alpha,\gamma}(t)]$ diverges because
the probability density $p^+_0(z)$ gives no finite moments. Thus,
the process $B[Z^U_{\alpha,\gamma}(t)]$ is subdiffusion, and
$B[Z^O_{\alpha,\gamma}(t)]$ is superdiffusion. In Fig.~\ref{fig2},
as an example, the propagator $p^r(x, t)$ for the under- and
overshooting anomalous diffusion with $\alpha=2/3$ and
$\gamma=2/3$ is drawn.

\begin{figure}
\centering
\includegraphics[width=8.6 cm]{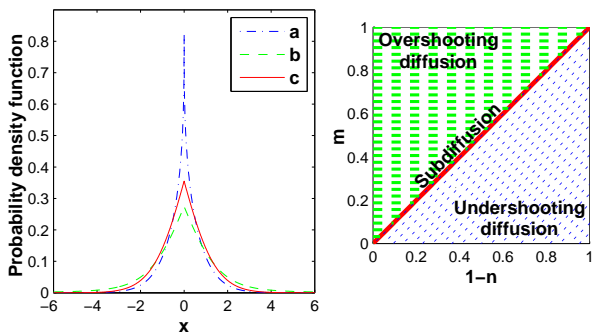}
\caption{(Color online) Left panel: the propagator $p^r(x, t)$ of
under- (a) and overshooting (b) anomalous diffusion with
$\alpha=2/3$ and $\gamma=2/3$ for $t=1$. The line (c) displays the
propagator of ordinary subdiffusion with $\alpha=2/3$ and
$\gamma=1$ for $t=1$. Right panel: diagram shows the interrelation
between $B[Z^U_{\alpha,\gamma}(t)]$, $B[S_\alpha(t)]$ and
$B[Z^O_{\alpha,\gamma}(t)]$. Here $m$ and $1-n$ denote the
power-law exponents defined in formulas (\ref{eqi}) and
(\ref{eqii}).}\label{fig3}
\end{figure}

It should be noticed that the ordinary subdiffusion
$B[S_\alpha(t)]$ takes an intermediate place between the under-
and overshooting anomalous diffusion  $B[Z^U_{\alpha,\gamma}(t)]$
and $B[Z^O_{\alpha,\gamma}(t)]$. The feature is illustrated in
Fig.~\ref{fig3}. This allows one to compare an asymptotic behavior
of the temporal evolution of diffusion fronts. From that one can
see that the diffusion front of $B[Z^U_{\alpha,\gamma}(t)]$ is
more stretched than the front of $B[S_\alpha(t)]$, whereas the
diffusion front of $B[Z^O_{\alpha,\gamma}(t)]$ is more contracted
in comparison with the front of $B[S_\alpha(t)]$.

One of interesting questions is what interpretation can be
assigned to the subordinators
$Z^U_{\alpha,\gamma}(t)=X^U_\gamma[S_\alpha(t)]$ and
$Z^O_{\alpha,\gamma}(t)=X^O_\gamma[S_\alpha(t)]$. As the processes
$X^U_\gamma(\tau)$ and $X^O_\gamma(\tau)$ are independent on
$S_\alpha(t)$, they can be considered separately. The inverse
L\'evy-stable process  $S_\alpha(t)$ accounts for the amount of
time, when a walker does not participate in motion. The pdf of the
subordinated process $B[X^U_\gamma(\tau)]$ is a special case of
the Dirichlet average, namely
\begin{displaymath}
F(\gamma,x,\tau)=\frac{\sin\pi\alpha}{\pi}\int_0^1p^B(x,\tau z)\,z^{\gamma-1}\,
(1-z)^{-\gamma}\,dz\,.
\end{displaymath}
Recall that many of important special and elementary functions can
be represented as Dirichlet averages of continuous functions (see
more details in \cite{carlson}). The Dirichlet average includes
the well-known means (arithmetic, geometric and others) as special
cases. The process $X^U_\gamma(t)$ evolves to infinity like time
$t$. Its contribution in the subordinated process
$B[X^U_\gamma(t)]$ is taken into account by  the Dirichlet average
of the probability density of the parent process $B$. The similar
reasoning can be developed for the process $X^O_\gamma(t)$.

\section{Conclusions}\label{par6}
The paper introduces an approach to study of the coupling between
the very large jumps in physical and operational times. It is
based on the compound subordination of a L\'evy-stable process
$T(\tau)$ by its inverse $S(t)$. The inverse L\'evy-stable process
is actually the left-inverse process of the L\'evy-stable process.
In fact, we have $S[T(\tau)]=\tau$, while $T[S(t)]>t$ holds. In
the framework of CTRWs and the Langevin-type stochastic
differential equations the compound subordinator provides a direct
coupling of physical and operational times. The subordination
scenario leads to two types of operational time: the spent
life-time and the residual age. In the first random process all
the moments are finite, whereas the second process has no finite
moments. We have shown that the approach is useful for analysis of
anomalous diffusion underlying all empirical fractional
two-power-law relaxation responses. Due to the two types of the
operational time the diffusion can display as well the
subdiffusive and superdiffusive character.

\section*{Acknowledgments}
AS is grateful to the Institute of Physics and the Hugo Steinhaus
Center for pleasant hospitality during his visit in Wroc{\l}aw
University of Technology. The authors also thank Dr. Marcin
Magdziarz for his remark to this work.


\end{document}